\documentstyle[12pt]{article}

\newcommand{\be}{\begin{equation}}
\newcommand{\ee}{\end{equation}}

\begin{document}

\begin{titlepage}

\begin{LARGE}
  \vspace{1cm}
  \begin{center}
     {The elliptic genus of Calabi-Yau 3- and 4-folds,
      product formulae and generalized Kac-Moody algebras.}
  \end{center}
\end{LARGE}

\vspace{15mm}

\begin{center}

  C.D.D. Neumann

  \vspace{2mm}

  {\it Department of Mathematics}\\
  {\it University of Amsterdam, }\\
  {\it Plantage Muidergracht 24, Amsterdam}\\
  {\it Email: {\tt neumann@fwi.uva.nl}}\\
  \vspace{40mm}

  \begin{large} ABSTRACT \end{large}

  \par

\end{center}
\begin{normalsize}
In this paper the elliptic genus for a general
Calabi-Yau fourfold is derived. The recent work 
of Kawai calculating N=2 heterotic string one-loop threshold
corrections with a Wilson line turned on is extended to
a similar computation where $K3$ is replaced by a general
Calabi-Yau 3- or 4-fold. In all cases there seems to be
a generalized Kac-Moody algebra involved, whose
denominator formula appears in the result.
\end{normalsize}

\end{titlepage}
\eject

\section{Introduction.}

In this paper, I extend the work of Kawai \cite{kawai},
calculating N=2 heterotic string one-loop threshold corrections with
a Wilson line turned on, to Calabi-Yau three- and fourfolds. 
(See also \cite{dijkg} for an alternative interpretation of
Kawai's result.) In
full generality, this calculation provides a map from
a certain class of Jacobi functions (including elliptic genera)
to modular functions of certain subgroups of
$Sp_4({\bf Q})$, in a product form. In a number of cases,
these products turn out to be equal to the denominator
formula of a generalized Kac-Moody algebra. It seems natural
to assume that this algebra is present in the corresponding string
theory, and indeed in \cite{BPS} it is argued that this
algebra is formed by the vertex operators of vector multiplets
and hypermultiplets.  

\section{Elliptic genus.}

In this section, I recall some basic facts about
elliptic genera for Calabi-Yau manifolds, mostly from \cite{elliptic},
and I explicitly derive it for 4-folds.
Let $C$ be a complex manifold of complex dimension $d$,
with $SU(d)$ holonomy. Then its elliptic genus is
a function $\phi(\tau,z)$ with the following
transformation properties
\be \phi \left(\frac{a\tau+b}{c\tau+d},\frac{z}{c\tau+d}\right)
  = e\left[ \frac{mcz^2}{c\tau+d} \right] \phi(\tau,z),
  \mbox{\hspace{1cm}} \left( \begin{array}{cc} a&b\\c&d
  \end{array} \right) \in SL_2({\bf Z}) \label{trans1} \ee
\be \phi(\tau,z+\lambda\tau+\mu)=(-1)^{2m(\lambda+\mu)}
    e[-m(\lambda^2\tau+2\lambda z)]
    \phi(\tau,z), \mbox{\hspace{5mm}} \lambda,\mu\in{\bf Z}
    \label{trans2}  \ee
where $m=\frac{d}{2}$, and it has an expansion of the
form
\be \phi(\tau,z) = \sum_{n\geq 0,r\in{\bf Z}+m}
  c(n,r)q^ny^r \ee
I use here the notations $e[x]=e^{2\pi ix}$, $q=e[\tau]$,
$y=e[z]$. The coefficients $c(0,-m+p)$ for $0\leq p\leq c$ have the
following geometrical meaning
\be c(0,-m+p) = \chi_p = \sum_{q=0}^{c}(-1)^{p+q} h^{p,q} \ee
where $h^{p,q}$ are the Hodge-numbers of $C$.
Furthermore
\be \phi(\tau,0) = \chi \ee
is the Euler number of $C$.
An important feature is that the elliptic genus can be
decomposed as
\be \phi(\tau,z) = \sum_{\mu=-m+1}^{m} h_{\mu}(\tau)
    \theta_{m,\mu}(\tau,z) \label{split} \ee
for functions $h_{\mu}$ and $\theta_{m,\mu}$ defined by
\be h_{\mu}(\tau) = \sum_{N \equiv -\mu^2 (\bmod 4m)}
    c_{\mu}(N) q^{N/4m} \ee
\be \theta_{m,\mu}(\tau,z) = \sum_{r \equiv \mu (\bmod 2m)}
    (-1)^{r-\mu}q^{r^2/4m} y^r \ee
Note that the $c_\mu(N)$ are only defined for
$-m+1\leq\mu\leq m$, but since $\theta_{m,\mu+2m}=(-1)^{2m}
\theta_{m,\mu}$, it is useful to define
\be c_r(N) = (-)^{r-\mu} c_\mu(N) \ee
for all $r\equiv\mu\bmod 2m$. The relation
between the coefficients of $h_\mu$ and $\phi$ is then given by
\be c(n,r)=c_r(4mn-r^2) \ee
Finally, the transformation properties of the $h_\mu$
can be derived to be
\be h_\mu(\tau+1) = e\left[-\frac{\mu^2}{4m}\right] h_\mu(\tau) \ee
\be h_\mu(-1/\tau) = \sqrt{i/2m\tau}\sum_{\nu=-m+1}^{m}
    e\left[\frac{\mu\nu}{2m}\right] h_\nu(\tau) \label{htran} \ee
Now if $m$ is integer, the elliptic genus satisfies
the defining properties of what is called a weak Jacobi
form of index $m$ and weight $0$. The ring $J_{2*,*}$
of weak Jacobi forms of even weight and all indices is
well known \cite{EZ}. It is a polynomial algebra over
$M_*$ (the ring of ordinary modular forms) with two
generators
\be A=\frac{\phi_{10,1}(\tau,z)}{\Delta(\tau)}, \mbox{\hspace{1cm}}
    B=\frac{\phi_{12,1}(\tau,z)}{\Delta(\tau)} \ee
Here $\Delta(\tau)=\eta^{24}(\tau)$ and $\phi_{10,1}$
and $\phi_{12,1}$ are unique cusp forms of index $1$
and weights $10$ and $12$ respectively. The generators
have an expansion
\be A = y^{-1} -2 +y +O(q) \ee
\be B = y^{-1}+10 +y +O(q) \ee
It immediately follows that the space $J_{0,1}$
is one dimensional with basis $B$, which implies
that the elliptic genus of a Calabi-Yau 2-fold is
\be \frac{\chi}{12} B \ee
So for $K3$, with $\chi=24$, it should be $2B$,
which is indeed the case \cite{K3}. The space $J_{0,2}$ is two
dimensional, with basis $E_4(\tau)A^2$ and $B^2$,
$E_4(\tau)$ being the normalized Eisenstein series of weight $4$.
So the elliptic genus is fixed by specifying
$\chi_0$ and $\chi$, leading to
\be \chi_0 E_4A^2 + \frac{\chi}{144}(B^2-E_4A^2) \ee
In the case that the manifold has strict $SU(d)$ holonomy,
which implies that $\chi_0=2$ the following
predictions can be done 
\be \chi_1 = 8-\frac{\chi}{6} \ee
\be \chi_2 = 12 + \frac{2\chi}{3} \ee
so that $\chi$ should be a multiple of six, and
there is a non trivial relation on the Hodge numbers
\be 4(h^{1,1}+h^{3,1})+44 = 2h^{2,1}+h^{2,2} \ee
as was recently noticed by Sethi, Vafa and Witten \cite{svw}.
For a Calabi-Yau 3-fold, the elliptic genus is known
to be \cite{elliptic}
\be \frac{\chi}{2}(y^{-\frac{1}{2}}+y^{\frac{1}{2}})
  \prod_{n=1}^{\infty}
  \frac{ (1-q^ny^2)(1-q^ny^{-2})}
  {(1-q^ny)(1-q^ny^{-1})} \label{cy3} \ee

\section{Product formulae.}

In this section
I look at the following generalization of
the formulae in \cite{kawai}
\be Z = \sum_{\mu=-m+1}^{m} Z_{m,\mu}(T,U,V,\tau) h_{\mu}(\tau) \ee
where the $h_\mu$ come from a function $\phi$, satisfying
the transformation properties~(\ref{trans1}) and~(\ref{trans2}),
and can be split like~(\ref{split}). For generality, I
allow this function 
to have a pole of finite order $N$ for $\tau \rightarrow i\infty$,
but nowhere else in the fundamental domain. So the function $\phi$
has a Fourier expansion of the form
\be \phi(\tau,z) = \sum_{n\geq-N, r\in{\bf Z}+m}
   c(n,r)q^n y^r \ee
converging for all $\tau$ with $\tau_2>0$ ($\tau_2 = \Im\tau$).
The functions $Z_{m,\mu}$ are defined by
\be Z_{m,\mu}(T,U,V,\tau) = \sum_{m_1,m_2,n_1,n_2}
    \sum_{b\in 2m{\bf Z}+\mu} (-1)^{b-\mu} q^{\frac{1}{2}p_L^2}
    \bar{q}^{\frac{1}{2}p_R^2} \ee
\be \frac{1}{2}p_R^2 = \frac{1}{4Y} |m_1U+m_2+n_1T+n_2(TU-mV^2)
    +bV|^2 \ee
\be \frac{1}{2}(p_L^2-p_R^2)=\frac{b^2}{4m}-m_1n_1+m_2n_2 \ee
\be Y = T_2U_2-mV_2^2 \ee
The function $Z$ is manifestly invariant under the following
transformations
\be U \rightarrow U+2\lambda mV+\lambda^2 mT,
    V \rightarrow V+\lambda T+\mu \ee
with $\lambda,\mu\in{\bf Z}$ if $m\in{\bf Z}$, and
$\lambda,\mu\in 2{\bf Z}$ if $m\in{\bf Z}+\frac{1}{2}$.
(This has the same effect as the substitutions
\[ m_2 \rightarrow m_2-\mu^2mn_2+\mu b \]
\be n_1 \rightarrow n_1+\lambda^2mm_1-2\lambda\mu mn_2 +\lambda b \ee
\[ b \rightarrow b+2\lambda mm_1 -2\mu mn_2 \]
and these leave the inproduct $\frac{b^2}{4m}-m_1n_1+m_2n_2$
invariant. In the same way one proves the other invariances.)
It is also invariant under the generalization of
$SL(2,{\bf Z})_T \times SL(2,{\bf Z})_U$, generated by
\be T\rightarrow T+1 \ee
\be T\rightarrow -\frac{1}{T}, U \rightarrow U-m\frac{V^2}{T},
    V\rightarrow \frac{V}{T} \ee
\be U\rightarrow U+1 \ee
\be U\rightarrow -\frac{1}{U}, T \rightarrow T-m\frac{V^2}{U},
    V\rightarrow \frac{V}{U} \ee
Furthermore, it is invariant under exchange of $T$ and $U$,
and under a parity transformation
\be T \leftrightarrow U \ee
\be V \rightarrow -V \ee
These transformations generate a group isomorphic
to $Sp_4({\bf Z})$ if $m=1$, and to a paramodular subgroup
of $Sp_4({\bf Q})$ \cite{grit} for $m>1$. Since
$\tau_2 Z$ is invariant under modular transformation
of $\tau$, as will be shown later,
the following integral is well defined and can
be evaluated explicitly by the methods of \cite{dixon,BPS}
\be {\cal I} = \int_{\cal F} \frac{d^2\tau}{\tau_2} (Z-c(0,0)) \ee
The subtraction is to remove the logarithmic singularities
due to the massless hypermultiplets, and is needed only if
$m$ is integer. If it is not, I define $c(0,0)$ to be zero.
Poisson resummation on $m_1,m_2$ leads to
\be \sum_{m_1,m_2} q^{\frac{1}{2}p_L^2} \bar{q}^{\frac{1}{2}p_R^2} =
    \sum_{k_1,k_2} \frac{Y}{U_2\tau_2}q^{\frac{1}{4m}b^2} \exp {\cal G} \ee
where
\[ {\cal G} = \frac{-\pi Y}{U_2^2 \tau_2} |A|^2
              -2\pi iT(n_1k_2+n_2k_1)
              +\frac{\pi b}{U_2} (V\tilde{A}-\bar{V}A) \]
\be           -\frac{\pi m n_2}{U_2} (V^2\tilde{A}-\bar{V}^2 A)
              +\frac{2\pi im V_2^2}{U_2^2} (n_1+n_2\bar{U}) A \ee
\be A = -k_1+n_1\tau +k_2U+n_2\tau U \ee
\be \tilde{A} = -k_1+n_1\tau +k_2\bar{U}+n_2\tau \bar{U} \ee
By applying another Poisson resummation on $b$,
it is easy to find the following transformation
properties of $Z_{\mu,m}$
\be Z_{\mu,m}(-1/\tau) = \sqrt{\tau/2mi} \sum_{\nu=-m}^{m}
    e\left[\frac{-\mu\nu}{2m}\right] Z_{\nu,m}(\tau) \ee
which together with the known properties~(\ref{htran})
of the $h_{\mu}$ prove the modular invariance of $\tau_2 Z$.
Following \cite{dixon,BPS} a bit further I find
\be {\cal I}_0 = \frac{Y}{U_2}\int \frac{d^2\tau}{\tau_2^2} \phi(\tau,0)
   = \frac{\pi}{3} E_2(\tau) \phi(\tau,0) |_{q^0} \ee
\be {\cal I}_d = \sum_{b\in {\bf Z}+m} 2\pi c(0,b)\left[
   b^2\frac{V_2^2}{U_2} - |b|V_2 + \frac{U_2}{6} \right]
   -c(0,0) \ln(kY) \ee
\[ -\ln \prod_{(l>0,b\in{\bf Z+m}),(l=0,0<b\in{\bf Z}+m)}
   |1-e[lU+bV]|^{4c(0,b)} \]
(This under the assumption that $0\leq V_2 \leq U_2/|b|$ for all
$b$ with $c(0,b)\neq 0$). Here
\be k=\frac{8\pi}{3\sqrt{3}} e^{1-\gamma} \ee
\be {\cal I}_{nd} = -\ln \prod_{k>0,l\in{\bf Z},b\in{\bf Z}+m}
  |1-e[kT+lU+bV]|^{4c(kl,b)} \ee
(This for $T_2$ large enough).
Putting this all together, I obtain
\be {\cal I} = -2\ln (kY)^{\frac{1}{2}c(0,0)}
    \left| e[pT+qU+rV] \prod_{(k,l,b)>0}
    (1-e[kT+lU+bV])^{c(kl,b)} \right|^2 \ee
where the coefficients $p,q,r$ are given by
\be p = \sum_{b\in{\bf Z}+m} \frac{b^2}{4m} c(0,b) \ee
\be q = \sum_{b\in{\bf Z}+m} \frac{1}{24} c(0,b) \ee
\be r = \sum_{b\in{\bf Z}+m} -\frac{|b|}{4} c(0,b) \ee
and the summation condition means $k>0$ or $k=0,l>0$ or
$k=l=0,b>0$ (always with $k,l\in{\bf Z}$ and $b\in{\bf Z}+m$).
Applying these formulae to $2B$, the elliptic
genus of $K3$, I recover the result of Kawai \cite{kawai}.
Now consider the elliptic genus of a Calabi-Yau fourfold,
\be \phi = \chi_0 E_4A^2 + \frac{\chi}{144}(B^2-E_4A^2) \ee
Amazingly, the $\chi$-dependent part equals
the coefficients of Gritsenko and Nikulin's second
product formula \cite{grit}, which is known to
be associated to the generalized Kac-Moody
algebra which is an automorphic form correction
to the Kac-Moody algebra defined by the symmetrized
generalized Cartan matrix
\be G_2=\left( \begin{array}{cccc}
4&-4&-12&-4 \\ -4&4&-4&-12 \\
-12&-4&4&-4 \\ -4&-12&-4&4 \end{array} \right) \ee
Unfortunately, there is no such formula for
the $\chi_0$-dependent part. So for a
Calabi-Yau fourfold I find
\be {\cal I} = -\chi_0 \ln((kY)^6|\Pi_6(\Omega)|^2)
  -\frac{\chi}{3}\ln((kY)^2 |F_2(\Omega)|^2) \ee
where $F_2$ is Gritsenko and Nikulins product and
$\Pi_6$ is
\be e[2V]\prod_{(k,l,b)>0} (1-e[kT+lU+bV])^{c(kl,b)} \ee
of weight $6$, with coefficients $c$ coming from $2E_4A^2$.
The following section describes the product
formula for a Calabi-Yau 3-fold. 

\section{Calabi-Yau 3-folds.}

In this section I apply my formulae to equation~(\ref{cy3}),
without the factor $\chi/2$. Expanding this in $q$ gives
\be (y^{-\frac{1}{2}}+y^{\frac{1}{2}}) +O(q) \ee
so that $c(0,-\frac{1}{2}) = c(0,\frac{1}{2}) = 1$,
and the corresponding product formula reads
\be F_0(T,U,V) = p^{\frac{1}{12}}q^{\frac{1}{12}}
  y^{-\frac{1}{4}} \prod_{(k,l,b)>0} (1-p^kq^ly^b)^{c(kl,b)} \ee
of weight zero,
where now $p=e[T]$, $q=e[U]$, $y=e[V]$. In the limit
$V\rightarrow 0$, this product behaves like
\be V\eta^2(p)\eta^2(q) \ee
as can be expected for $\chi=2$.
This product can be expanded in terms of $p$ (since
it is valid for $T_2$ large enough). It turns out
to be useful to consider $F_0(T,U,2V)$. Thus
\be F_0(T,U,2V) = \sum_{k\in{\bf Z}_{\geq 0}+\frac{1}{12}}
   \phi_k(q,y) p^k \ee
This is a variant of what is known as a Fourier-Jacobi 
expansion. The transformation properties of $F_0(T,U,V)$ 
imply that the coefficients $\phi_m$ should be
Jacobi forms of weight $0$ and index $6k$, with
a possible multiplier system. From the product
formula it is possible to read of the lowest order
coefficient
\be \phi_{\frac{1}{12}}(q,y) = q^{\frac{1}{12}}
  ( y^{-\frac{1}{2}} - y^{\frac{1}{2}})
   \prod_{n>0}(1-q^ny)(1-q^ny^{-1})
  = \theta_{11}(q,y)\eta^{-1}(q) \ee
by the product formula for theta-functions. This is indeed
a Jacobi cusp form of weight $0$ and index $\frac{1}{2}$
with multiplier system \cite{mumford}, which can serve
as a consistency check. It can be written
as a sum as follows
\be \left( \sum_{n\in {\bf Z}} (-1)^n q^{\frac{(2n+1)^2}{8}}
    y^{\frac{(2n+1)}{2}} \right) \left( \sum_{n\geq 0}
    p(n) q^{n-\frac{1}{24}} \right) \ee
where $p(n)$ is the partition function. 
Now unlike the case
of $F_2(\Omega)$ from \cite{grit}, is doesn't seem
to be possible to write the entire product as a lifting of
its first Fourier-Jacobi coefficient.
It does seem to be likely that this function
is also related to some generalized Kac-Moody algebra.
This is under investigation. The final result for the
Calabi-Yau 3-fold calculation is
\be {\cal I} = -\chi\ln|F_0(\Omega)|^2 \ee

\vspace{1\baselineskip} \noindent
{\bf Acknowledgement.} 

\noindent I would like to thank
R. Dijkgraaf for helpful discussions.

\end{document}